\documentstyle[12pt,psfig]{article}

\catcode`\@=11

\def\NPB#1#2#3{{\it Nucl.~Phys.} {\bf{B#1}} (19#2) #3}
\def\PLB#1#2#3{{\it Phys.~Lett.} {\bf{B#1}} (19#2) #3}
\def\PRD#1#2#3{{\it Phys.~Rev.} {\bf{D#1}} (19#2) #3}

\def\AP#1#2#3{{\it Ann.~Phys.} {\bf#1} (19#2) #3}

\def\@normalsize{\@setsize\normalsize{15pt}\xiipt\@xiipt
\abovedisplayskip 14pt plus3pt minus3pt%
\belowdisplayskip \abovedisplayskip
\abovedisplayshortskip  \z@ plus3pt%
\belowdisplayshortskip  7pt plus3.5pt minus0pt}
\def\small{\@setsize\small{13.6pt}\xipt\@xipt
\abovedisplayskip 13pt plus3pt minus3pt%
\belowdisplayskip \abovedisplayskip
\abovedisplayshortskip  \z@ plus3pt%
\belowdisplayshortskip  7pt plus3.5pt minus0pt
\def\@listi{\parsep 4.5pt plus 2pt minus 1pt
            \itemsep \parsep
            \topsep 9pt plus 3pt minus 3pt}}

\def\underline#1{\relax\ifmmode\@@underline#1\else
        $\@@underline{\hbox{#1}}$\relax\fi}
\@twosidetrue
\relax

\catcode`@=12

\catcode`\@=11

\def\section{\@startsection{section}{1}{\z@}{3.5ex plus 1ex minus
   .2ex}{2.3ex plus .2ex}{\large\bf}}

\def\ps@headings{\def\@oddfoot{}\def\@evenfoot{}
\def\@oddhead{\hbox{}\hfill
        \makebox[.5\textwidth]{\raggedright\ignorespaces --\thepage{}--
        \hfill }}
\def\@evenhead{\@oddhead}
\def\subsectionmark##1{\markboth{##1}{}}
}

\ps@headings

\catcode`\@=12

\relax

\def\figcap{\section*{Figure Captions\markboth
        {FIGURECAPTIONS}{FIGURECAPTIONS}}\list
        {Fig. \arabic{enumi}:\hfill}{\settowidth\labelwidth{Fig. 999:}
        \leftmargin\labelwidth
        \advance\leftmargin\labelsep\usecounter{enumi}}}
 \relax
\def\tablecap{\section*{Table Captions\markboth
        {TABLECAPTIONS}{TABLECAPTIONS}}\list
        {Table \arabic{enumi}:\hfill}{\settowidth\labelwidth{Table 999:}
        \leftmargin\labelwidth
        \advance\leftmargin\labelsep\usecounter{enumi}}}
 \relax
\def\reflist{\section*{References\markboth
        {REFLIST}{REFLIST}}\list
        {[\arabic{enumi}]\hfill}{\settowidth\labelwidth{[999]}
        \leftmargin\labelwidth
        \advance\leftmargin\labelsep\usecounter{enumi}}}
 \relax

\catcode`\@=11

\def\marginnote#1{}
\newcount\hour
\newcount\minute
\newtoks\amorpm
\hour=\time\divide\hour by60
\minute=\time{\multiply\hour by60 \global\advance\minute by-
\hour}
\edef\standardtime{{\ifnum\hour<12 \global\amorpm={am}%
    \else\global\amorpm={pm}\advance\hour by-12 \fi
    \ifnum\hour=0 \hour=12 \fi
    \number\hour:\ifnum\minute<100\fi\number\minute\the\amorpm}}
\edef\militarytime{\number\hour:\ifnum\minute<100\fi\number\minute}
\def\draftlabel#1{{\@bsphack\if@filesw {\let\thepage\relax
  \xdef\@gtempa{\write\@auxout{\string
    \newlabel{#1}{{\@currentlabel}{\thepage}}}}}\@gtempa
    \if@nobreak \ifvmode\nobreak\fi\fi\fi\@esphack}
     \gdef\@eqnlabel{#1}}
\def\@eqnlabel{}
\def\@vacuum{}
\def\draftmarginnote#1{\marginpar{\raggedright\scriptsize\tt#1}}
\def\draft{\oddsidemargin -.5truein
        \def\@oddfoot{\sl preliminary draft \hfil
        \rm\thepage\hfil\sl\today\quad\militarytime}
        \let\@evenfoot\@oddfoot \overfullrule 3pt
        \let\label=\draftlabel
        \let\marginnote=\draftmarginnote
   
\def\@eqnnum{(\theequation)\rlap{\kern\marginparsep\tt\@eqnlabel}%
\global\let\@eqnlabel\@vacuum}  }
\def\preprint{\twocolumn\sloppy\flushbottom\parindent 1em
        \leftmargini 2em\leftmarginv .5em\leftmarginvi .5em
        \oddsidemargin -.5in    \evensidemargin -.5in
        \columnsep 15mm \footheight 0pt
        \textwidth 250mmin      \topmargin  -.4in
        \headheight 12pt \topskip .4in
        \textheight 175mm
        \footskip 0pt
        
\def\@oddhead{\thepage\hfil\addtocounter{page}{1}\thepage}
        \let\@evenhead\@oddhead \def\@oddfoot{} \def\@evenfoot{} 
}
\def\titlepage{\@restonecolfalse\if@twocolumn\@restonecoltrue\onecolumn
     \else \newpage \fi \thispagestyle{empty}\c@page\z@
        \def\thefootnote{\fnsymbol{footnote}} }
\def\endtitlepage{\if@restonecol\twocolumn \else  \fi
        \def\thefootnote{\arabic{footnote}}
        \setcounter{footnote}{0}}  
\catcode`@=12
\relax

\def\ps@headings{\def\@oddfoot{}\def\@evenfoot{}
\def\@oddhead{\hbox{}\hfill
        \makebox[.5\textwidth]{\raggedright\ignorespaces --\thepage{}--
        \hfill }}
\def\@evenhead{\@oddhead}
\def\subsectionmark##1{\markboth{##1}{}}
}

\ps@headings

\relax

\def\firstpage#1#2#3#4#5#6{
\begin{document}
\begin{titlepage}
\nopagebreak
\title{\begin{flushright}
        \vspace*{-1.8in}
        {\normalsize CERN-TH/97-165}\\[-9mm]
        {\normalsize CPTH-S548.0797}\\[-9mm]
        {\normalsize IEM-FT-160/97}\\[-9mm]
        {\normalsize hep-th/9707208}\\[4mm]
\end{flushright}
\vfill
{#3}}
\author{\large #4 \\[1.0cm] #5}
\maketitle
\vskip -7mm     
\nopagebreak 
\begin{abstract}
{\noindent #6}
\end{abstract}
\vfill
\begin{flushleft}
\rule{16.1cm}{0.2mm}\\[-3mm]
$^{\star}${\small Research supported in part by\vspace{-4mm}
IN2P3-CICYT under contract Pth 96-3, in part by 
CICYT contract AEN95-0195, and
in part by the EEC under the TMR contract 
ERBFMRX-CT96-0090.}\\[-3mm] 
$^{\dagger}${\small Laboratoire Propre du CNRS UPR A.0014.}\\
CERN-TH/97-165\\
July 1997
\end{flushleft}
\thispagestyle{empty}
\end{titlepage}}

\def\simlt{\stackrel{<}{{}_\sim}}
\def\simgt{\stackrel{>}{{}_\sim}}
\newcommand{\dal}{\raisebox{0.085cm}
{\fbox{\rule{0cm}{0.07cm}\,}}}
\newcommand{\dt}{\partial_{\langle T\rangle}}
\newcommand{\dtbar}{\partial_{\langle\overline{T}\rangle}}
\newcommand{\al}{\alpha^{\prime}}
\newcommand{\mst}{M_{\scriptscriptstyle \!S}}
\newcommand{\mpl}{M_{\scriptscriptstyle \!P}}
\newcommand{\dv}{\int{\rm d}^4x\sqrt{g}}
\newcommand{\lv}{\left\langle}
\newcommand{\rv}{\right\rangle}
\newcommand{\ph}{\varphi}
\newcommand{\abar}{\overline{a}}
\newcommand{\sbar}{\,\overline{\! S}}
\newcommand{\xbar}{\,\overline{\! X}}
\newcommand{\fbar}{\,\overline{\! F}}
\newcommand{\zbar}{\overline{z}}
\newcommand{\dbar}{\,\overline{\!\partial}}
\newcommand{\tbar}{\overline{T}}
\newcommand{\taubar}{\overline{\tau}}
\newcommand{\ubar}{\overline{U}}
\newcommand{\ybar}{\overline{Y}}
\newcommand{\phb}{\overline{\varphi}}
\newcommand{\cm}{Commun.\ Math.\ Phys.~}
\newcommand{\prl}{Phys.\ Rev.\ Lett.~}
\newcommand{\pr}{Phys.\ Rev.\ D~}
\newcommand{\pl}{Phys.\ Lett.\ B~}
\newcommand{\ibar}{\overline{\imath}}
\newcommand{\jbar}{\overline{\jmath}}
\newcommand{\np}{Nucl.\ Phys.\ B~}
\newcommand{\F}{{\cal F}}
\renewcommand{\L}{{\cal L}}
\newcommand{\A}{{\cal A}}
\newcommand{\e}{{\rm e}}
\newcommand{\be}{\begin{equation}}
\newcommand{\en}{\end{equation}}
\newcommand{\ba}{\begin{eqnarray}}
\newcommand{\dslash}{{\not\!\partial}}
\newcommand{\ea}{\end{eqnarray}}
\newcommand{\gsi}{\,\raisebox{-0.13cm}{$\stackrel{\textstyle
>}{\textstyle\sim}$}\,}
\newcommand{\lsi}{\,\raisebox{-0.13cm}{$\stackrel{\textstyle
<}{\textstyle\sim}$}\,}
\date{}
\firstpage{3118}{IC/95/34}
{\large\bf On the M-theory description of gaugino condensation} 
{I. Antoniadis$^{\,a,b}$ and M. Quir{\'o}s$^{\,c}$}
{\normalsize\sl
$^a$TH-Division, CERN, CH-1211 Geneva 23, Switzerland\\[-3mm]
\normalsize\sl$^b$ Centre de Physique Th{\'e}orique, 
Ecole Polytechnique,$^\dagger$ {}F-91128 Palaiseau, France\\[-3mm]
\normalsize\sl
$^c$Instituto de Estructura de la Materia, CSIC, Serrano 123, 28006 Madrid,
Spain}
{We present an additional test of the recent proposal for describing 
supersymmetry breaking due to gaugino condensation in the strong coupling 
regime, by a Scherk-Schwarz mechanism on the eleventh dimension of M-theory. 
An analysis of supersymmetric transformations in the infinite-radius limit 
reveals the presence of a discontinuity in the spinorial parameter, which 
coincides with the result found in the presence of gaugino condensation. 
The condensate is then identified with the quantized parameter entering 
the modification of the Scherk-Schwarz boundary conditions. 
This mechanism provides an alternative perturbative explanation of the
gauge hierarchy that determines the scale of low-energy supersymmetry breaking
in terms of the unification gauge coupling.}

\newpage
The 10-dimensional (10D) $E_8\times E_8$ heterotic string, compactified on an
appropriate 6D internal manifold, is a good candidate for 
describing the observed low-energy world. In particular, compactification 
on a Calabi-Yau (CY) manifold leads to a 4D $N=1$ supersymmetric theory
that can accommodate the gauge group and matter content of the standard model.
On the other hand, there is a mismatch between the gauge coupling unification 
scale, $M_G\sim 10^{16}$ GeV, and the heterotic string scale $M_H$, which is
determined in terms of the Planck scale, $M_p\sim 10^{19}$ GeV, 
from the relation $M_H=(\alpha_G/8)^{1/2}M_p\sim 10^{18}$ GeV, where
$\alpha_G\sim 1/25$ is the unification gauge coupling. 
However, this relation
does not hold if the compactification scale is small compared to 
$M_H$, in which case the 10D theory is strongly interacting. 

It is now believed that the strong coupling limit of the 10D heterotic string
theory compactified on CY is described
by the 11D M-theory compactified on CY$\times S^1/{\bf Z}_2$
upon the identification~\cite{hw,w}:
\be
M_{11}\equiv 2\pi(4\pi\kappa_{11}^2)^{-1/9}
=2\pi(2\alpha_G {\widehat V})^{-1/6}\qquad ;\qquad
\rho^{-1}={4\over\alpha_G}M_{11}^3 M^{-2}_p\, ,
\label{rels}
\en
where $\kappa_{11}$ is the 11D gravitational coupling, $\rho$ is the radius
of the semicircle $S^1/{\bf Z}_2$, and ${\widehat V}$ is the CY volume.
In this regime, the value of the unification scale can become
consistent with the M-theory scale $M_{11}\sim M_G$, if the radius $\rho$ 
of the semicircle is at an intermediate scale
$\rho^{-1}\sim 10^{12}$ GeV, while for isotropic CY the compactification scale 
${\widehat V}^{-1/6}/2\pi$ is of the order of
$M_{11}$. Fortunately, this is inside the region of validity of M-theory, 
$\rho M_{11}\gg 1$ and ${\widehat V}\kappa_{11}^{-4/3}\gg 1$.
As a result, the effective theory above the intermediate scale behaves as
5-dimensional, but only in the gravitational and moduli sector; the gauge
sectors coming from $E_8\times E_8$ live at the 4D boundaries of the 
semicircle.

In recent works~\cite{aq,dg,aq2}, the
intermediate scale $\rho^{-1}$ was related with the scale of supersymmetry
breaking by means of a coordinate-dependent compactification of
the eleventh dimension, analogue to the Scherk-Schwarz
mechanism~\cite{ss}. The observable world living at the boundaries 
remains unaffected and
feels supersymmetry breaking only by gravitational interactions,
which yield $m_{\rm susy}\sim \rho^{-2}/M_p$~\cite{aq,aq2}. Moreover we
suggested~\cite{aq2} that this (perturbative) mechanism describes
ordinary (non-perturbative) gaugino condensation~\cite{gc} 
in the strongly coupled heterotic string. In particular,
assuming that the basic relations of gaugino condensation in the
weakly coupled heterotic string, $m_{3/2}=|W|e^{K/2}\propto\Lambda_c^3$, 
also hold in the strong coupling regime, and using 
the fact that the superpotential $W$
is of order~1 (in Planck units) at the minimum, 
we found simple scaling relations with the volume 
${\widehat V}\sim e^{-K}$: $m_{3/2}\sim {\widehat V}^{-1/2}$ and 
$\Lambda_c\sim{\widehat V}^{-1/6}$. Comparison with the duality relations
(\ref{rels}) yields the following identifications for the gravitino mass 
and the condensation scale: $m_{3/2}\sim\rho^{-1}$ and 
$\Lambda_c\sim M_{11}$~\cite{aq2}, respectively.

In this letter we further analyse the mechanism of supersymmetry
breaking in M-theory by Scherk-Schwarz compactification on the 
eleventh dimension and present additional evidence that
it provides a dual description of gaugino
condensation in the strongly coupled heterotic string. 
We find that the goldstino is the (right-handed) 
fifth component of the 5D gravitino, while there is a
discontinuity in the supersymmetric spinorial transformation
parameter around the end-point $\pm\pi\rho$ of the semicircle.
This discontinuity survives in the limit $\rho\rightarrow \infty$, where 
the gravitino mass vanishes and supersymmetry is locally restored, in agreement
with the result previously found in the case of gaugino condensation in the 
strongly coupled heterotic string~\cite{h}. 
Furthermore, the quantization of the
condensate is related to the quantized parameter entering the
Scherk-Schwarz boundary conditions. 
Finally, consistency of the proposed dual description
requires that the hidden $E_8$ be strongly coupled at the M-theory scale, which
determines $\rho^{-1}$ in terms of $\alpha_G$ 
at the desired intermediate scale $\sim 10^{12}$ GeV. 
Consequently, the large hierarchy between the supersymmetry-breaking scale in 
the observable sector and the Planck mass arises as a result of 
successive power
suppressions of the gauge coupling of the form $(\alpha_G/16\pi^2)^4$.

We start by first considering the $N=1$ 5D theory obtained from
compactification of M-theory on a CY manifold with Hodge numbers
$h_{(1,1)}$ and $h_{(1,2)}$~\cite{ccaf}. In addition to the
gravitational multiplet, the massless spectrum contains
$n_V=h_{(1,1)}-1$ vector multiplets and $n_H=h_{(1,2)}+1$
hypermultiplets. 
The $N=1$ supersymmetry transformations in the 5D theory are \cite{sierra}:
\ba
\delta e_M^m &=& -{i\over 2}{\overline{\cal E}}\Gamma^m\Psi_M
\nonumber\\
\delta \Psi_M &=& D_M{\cal E} +\cdots
\nonumber\\
\delta{\cal X}^a &=& -{1\over 2} v^a_\alpha(\dslash\phi^\alpha)
{\cal E} +\cdots\nonumber\\
\delta\phi^\alpha &=& {i\over 2} v_a^\alpha{\overline{\cal E}}{\cal X}^a\, ,
\label{susy}
\ea
where $e_M^m$ is the f\"unfbein, $\Gamma^m=(\gamma^\mu,i\gamma_5)$ are the
Dirac matrices~\footnote{The $\Gamma^m$ matrices are defined by
their anticommutation rules, $\left\{\Gamma^m,\Gamma^n\right\}=2\eta^{mn}$,
where the space-time metric is $\eta^{mn}={\rm diag}(1,-1,-1,-1,-1)$. They
satisfy the relation $\Gamma^0\cdots\Gamma^4=1$. The 4D matrices
$\Gamma^\mu=\gamma^\mu$ ($\mu=0,\dots ,3$) are purely imaginary and
$\Gamma^4=i\gamma_5$ is real.}, $\Psi_M$ is the gravitino field, ${\cal E}$ the
spinorial parameter of the transformation, and the dots stand for non-linear
terms. Finally, ${\cal X}^a$ and $\phi^\alpha$ denote the fermionic and scalar
components of vector multiplets while $v^a_\alpha$ is the vielbein of the
corresponding moduli space. Similar transformations hold for the components of
hypermultiplets for which our subsequent analysis can be generalized in a
straightforward way. 

All fermions in eq.~(\ref{susy}) can be represented as doublets under the 
$SU(2)$ $R$-symmetry whose components are subject to the (generalized) Majorana
condition; in a suitable basis~\cite{cremmer}:
\be
\Psi\equiv\left(
\begin{array}{c}\psi_1\\\psi_2\end{array}\right)
=\gamma_5\left(
\begin{array}{c}\psi_2^*\\-\psi_1^*\end{array}\right)\, ,
\label{doublet}
\en
where $\Psi$ describes any generic (Dirac) spinor of 
eq.~(\ref{susy})~\footnote{Hereafter we will conventionally denote fermionic 
$SU(2)_R$ doublets with upper-case symbols and their corresponding
components with lower-case ones.}. In this notation, it is understood that the
$\gamma$-matrices act diagonally in the $SU(2)_R$ space. It is
convenient to decompose the spinors with respect to the four-dimensional
chirality. Using the relations $\gamma_5^2=1$ and 
$\gamma_5^*=-\gamma_5$, which are 
valid in the above basis, it follows that $\gamma_5\psi_1=\pm\psi_1$ implies 
$\gamma_5\psi_1^*=\mp\psi_1^*$. We can then define:
\be
\Psi_L\equiv\left(
\begin{array}{c}\psi_{L}\\\psi_{R}^*\end{array}\right)\qquad\qquad
\Psi_R\equiv\left(
\begin{array}{c}\psi_{R}\\-\psi_{L}^*\end{array}\right)\, ,
\label{LR}
\en
in terms of the 4D chiral spinors $\psi_{L,R}=\pm\gamma_5\psi_{L,R}\, ,$ where
$\psi^*_{R,L}\equiv(\psi_{L,R})^*$. This decomposition amounts, 
in terms of $SU(2)_R$ doublets, to the condition
\be
\Gamma_5\Psi_{L,R}=\pm\Psi_{L,R}\qquad ;\qquad
\Gamma_5=\left(
\begin{array}{cc}\gamma_5&0\\0&-\gamma_5\end{array}\right)\ .
\label{chiral}
\en
The ``chiral" spinors $\Psi_{L,R}$
satisfy trivially the condition (\ref{doublet}). 
Furthermore, it is easy to show the relations 
$${\overline{\cal E}}_L(\gamma^\mu)^{2n}\Psi_R=2i{\rm Im}\,
\left\{\overline\varepsilon_L
(\gamma^\mu)^{2n}\psi_R\right\},\qquad
{\overline{\cal E}}_L(\gamma^\mu)^{2n+1}\Psi_L=2i{\rm Im}\,
\left\{\overline\varepsilon_L(\gamma^\mu)^{2n+1}\psi_L\right\},$$ 
$${\overline{\cal E}}_L(\gamma^\mu)^{2n}\gamma_5\Psi_R=2{\rm Re}\,
\left\{\overline\varepsilon_L
(\gamma^\mu)^{2n}\gamma_5\psi_R\right\},\qquad
{\overline{\cal E}}_L(\gamma^\mu)^{2n+1}\gamma_5\Psi_L=2{\rm Re}\,
\left\{\overline\varepsilon_L(\gamma^\mu)^{2n+1}\gamma_5\psi_L\right\},$$ 
which are also valid when $L$ and $R$ are interchanged.

Upon compactification to $D=4$ on the semi-circle $S^1/{\bf Z}_2$ of radius
$\rho$, one obtains an $N=1$ supersymmetric theory that, for large $\rho$, 
describes the strong coupling limit of the heterotic
string compactified on the same Calabi-Yau manifold 
as compactifies the 11-dimensional
theory to $D=5$ \cite{hw,w}. The gauge group then appears 
at the two end-points of the semicircle and consists of a subgroup of 
$E_8\times E_8$ together with some matter representations depending on the
particular embedding of the spin connection. For instance, for the standard 
embedding into a
single $E_8$, one obtains an $E_6$ sitting at one end $(y=\pi\rho)$ 
together with $h_{(1,1)}$ ${\bf 27}$'s and $h_{(1,2)}$ ${\overline{\bf 27}}$'s, 
and a pure $E_8$ sitting at the other end $(y=0)$.

The ${\bf Z}_2$ projection is defined through the reflection ${\cal R}$ 
of the fifth
coordinate $y$ parametrizing $S^1$, and has the following action on the
5D fermionic fields:
\be
{\cal R}\Psi(x^\mu,y)\equiv\eta\Gamma_5\Psi(x^\mu,-y)\, ,
\label{R}
\en
with $\eta=1$ for $\Psi=\Psi_\mu$ and $\eta=-1$ for $\Psi=\Psi_5,{\cal X}$.
In terms of the spinors defined in eq.~(\ref{LR}), 
\be
{\cal R}\Psi_{L,R}(x^\mu,y)=\pm\eta\Psi_{L,R}(x^\mu,-y)\ . 
\label{Rchiral}
\en
The 
${\bf Z}_2$ projection is defined by keeping the states that are even under
$\cal R$. It follows that the remaining massless fermions are the left-handed
components of the 4D gravitino $\Psi_{\mu L}$, as well as the
right-handed components of $\Psi_{5R}$ and ${\cal X}_R$. 
Taking into account the 
${\bf Z}_2$ action in the bosonic sector, which projects away the off-diagonal
components of the f\"unfbein ($e_\mu^5$), the above massless spectrum is 
consistent with the residual $N=1$ supersymmetry transformations at $D=4$
given by eq.~(\ref{susy}) with a fermionic parameter $\cal E$ reduced
to its left-handed component ${\cal E}_L$.

In order to spontaneously break supersymmetry,
we apply the Scherk-Schwarz mechanism on the fifth coordinate $y$ \cite{ss}. 
For this purpose, we need an $R$-symmetry, which transforms 
the gravitino non-trivially and imposes boundary conditions, around $S^1$, 
which are periodic up to a symmetry transformation:
\be
\Psi_M(x^\mu,y+2\pi\rho)=e^{2i\pi\omega Q}\Psi_M(x^\mu,y)\, ,
\label{bc}
\en
where $Q$ is the $R$-symmetry generator and $\omega$ the transformation
parameter. The continuous symmetry is in general broken by the compactification
to some discrete subgroup, leading to quantized values of $\omega$. 
For instance,
in the case of ${\bf Z}_N$ one has $\omega=1/N$ and $Q=0,\dots,N-1$.
For generic values of $\omega$,
eq.~(\ref{bc}) implies that the zero mode of the gravitino acquires
an explicit $y$-dependence:
\be
\Psi_M(x^\mu,y)=U(y)\Psi^{(0)}_M(x^\mu)+\cdots\quad ;\quad 
{\displaystyle U=e^{i{\omega\over\rho}y Q}}\, ,
\label{mode}
\en
where the dots stand for Kaluza-Klein (KK) modes.

Consistency of the theory requires that the matrix $U$ commutes with the
reflection $\cal R$, which defines the $N=1$ projection~\cite{largeR,modelsR}.
{}From eq.~(\ref{R}) one then finds:
\be
\Gamma_5 U(-y)=U(y) \Gamma_5\, ,
\label{cons}
\en
implying that the generator $Q$ anticommutes with $\Gamma_5$, 
$\{Q,\Gamma_5\}=0$~\footnote{Notice that condition (\ref{cons}) guarantees 
that the $\cal R$-chirality of a spinor, $\Psi_{L,R}(x^\mu,y)$, 
in the sense of eq.~(\ref{Rchiral}), coincides
with the $\Gamma_5$-chirality of its zero-mode $\Psi_{L,R}^{(0)}(x^\mu)$, in 
the sense of eq.~(\ref{chiral}). In this way one can write the decomposition
(\ref{mode}) for the chiral components of $\Psi$, i.e.
$\Psi_{L,R}(x^\mu,y)=U(y)\Psi_{L,R}^{(0)}(x^\mu)$.}
. The general solution is \cite{dg}:
\be
Q=\sin\theta\sigma_1+\cos\theta\sigma_2\quad ;\quad
U=\left(
\begin{array}{cc}\cos{\omega y\over\rho}&\sin{\omega y\over\rho}e^{i\theta}\\
-\sin{\omega y\over\rho}e^{-i\theta}&
\cos{\omega y\over\rho}\end{array}\right)\, ,
\label{U}
\en
where $\sigma_{1,2}$ are the Pauli matrices representing 
$SU(2)_R$ generators and $\theta$ is an arbitrary
(real) parameter. For the particular value $\omega=1/2$ there is an additional
solution to eqs.~(\ref{bc}) and (\ref{cons}) corresponding to~\cite{aq2}:
\be
Q=\left(
\begin{array}{cc}\theta_1&0\\0&\theta_2\end{array}\right)\quad ;\quad
U=\left(
\begin{array}{cc}e^{i\pi\theta_1}&0\\0&e^{i\pi\theta_2}\end{array}\right)
\cos{y\over 2\rho}\quad \quad
\left(\omega={1\over 2}\right)\, ,
\label{U12}
\en
with $\theta_{1,2}$ arbitrary constants. Note however that this solution 
involves both $n=0$ and $n=-1$ KK-modes, which makes the effective field 
theory description of the spontaneous supersymmetry breaking more complicated. 
{}For this reason, we restrict our subsequent analysis to solution (\ref{U}).

Inspection of the supersymmetry transformations (\ref{susy}), together with 
the requirement that the f\"unfbein zero mode does not depend on $y$, shows
that the $y$-dependence of the supersymmetry parameter is the same as that of
the gravitino zero-mode, 
i.e. ${\cal E}(x^\mu,y)=U(y){\cal E}^{(0)}(x^\mu)$
\cite{ss}. Supersymmetry in the 4D theory is then spontaneously broken, with 
the goldstino being identified with the fifth component of the 5D gravitino,
$\Psi_5^{(0)}$. Indeed, for global supersymmetry parameter,
$D_\mu{\cal E}^{(0)}=0$, its variation is~\footnote{Note that the operator
$U^{-1}\partial_y U$ turns a left-handed spinor, in the sense of
eq.~(\ref{chiral}), into a right-handed one.}:
\be
\delta\Psi_5^{(0)}=(U^{-1}\partial_yU){\cal E}^{(0)} +\cdots
\quad ;\quad
U^{-1}\partial_yU={\omega\over\rho}\left(
\begin{array}{cc}0&e^{i\theta}\\-e^{-i\theta}&0\end{array}\right)\, ,
\label{gold}
\en
while no other fermions can acquire finite constant shifts in their 
transformations. The reason is that the scalar components of the zero-mode
supermultiplets are either inert under the $R$-symmetry, and therefore
$y$-independent, or otherwise become massive, and $y$-dependent as 
$\phi(x^\mu,y)=v+U(y)\phi^{(0)}(x^\mu)+\cdots$, with $v$ being a constant and
the quantum fluctuation $\phi^{(0)}$ having zero vacuum expectation value (VEV). 
It is then clear from eq.~(\ref{susy})
that the variation of their fermionic superpartners vanishes in the
vacuum, $\delta{\cal X}^{(0)}=0$.

The kinetic term of the 5D gravitino,
$$-i/2 \overline{\Psi}_M\Gamma^{MNP}D_N\Psi_P,$$ 
where ${\displaystyle \Gamma^{M_1\dots M_n}\equiv\Gamma^{[M_1}
\Gamma^{M_2}\dots \Gamma^{M_n]} }$ is the totally antisymmetric product,
gives rise to the following 4-dimensional 
Lagrangian for the gravitino zero modes:
\be
e^{-1}{\cal L}^{(0)}=
-\frac{i}{2}\overline{\Psi}^{(0)}_M\Gamma^{MNP}D_N\Psi^{(0)}_P
-\frac{1}{2}\overline{\Psi}^{(0)}_\mu\left(U^{-1}\partial_y
U\right)\Gamma^{\mu\nu}\gamma_5\Psi^{(0)}_\nu\, ,
\label{lag0}
\en
where the matrix $U^{-1}\partial_y U$ is given in eq.~(\ref{gold}). The
first term in the r.h.s. of eq.~(\ref{lag0}) is the kinetic term of the 
gravitino zero modes, while
the second is a mass term for the $\Psi^{(0)}_\mu$ component. Notice
that the goldstino component $\Psi^{(0)}_5$ remains massless, as it should.

The above arguments are also valid in the $N=1$ theory, 
obtained by applying the
${\bf Z}_2$ projection defined through the $\cal R$-reflection (\ref{R}).
The $y$-dependence of the remaining zero modes is always given by
eq.~(\ref{mode}). The goldstino is now the right-handed component 
$\psi^{(0)}_{5R}$, which, from eq.~(\ref{gold}), transforms as:
\be
\delta\psi_{5R}^{(0)}={\omega\over\rho}e^{i\theta}\varepsilon^{(0)*}_R+\cdots
\label{goldR}
\en
The surviving gravitino is $\Psi^{(0)}_{\mu L}$ in the notation of
eq.~(\ref{LR}). Its kinetic term can be read off from 
the first term in the r.h.s. of eq.~(\ref{lag0}):
\be
e^{-1}{\cal L}^{(0)}_{\rm kin}=-\frac{i}{2}\overline{\Psi}^{(0)}_{\mu
L}\Gamma^{\mu\nu\rho}D_\nu \Psi^{(0)}_{\rho L}=
-\frac{i}{2}\overline{\psi}^{(0)}_{\mu
L}\Gamma^{\mu\nu\rho}D_\nu\psi^{(0)}_{\rho L}+\ h.c.
\label{kin}
\en
while the second term yields a mass for the gravitino zero mode 
equal to $\omega/\rho$:
\be
e^{-1}{\cal L}^{(0)}_m=-\frac{1}{2}\frac{\omega}{\rho}\
\overline{\Psi}^{(0)}_{\mu L}
\left(
\begin{array}{cc}
0 & e^{i\theta}\\
-e^{-i\theta} & 0
\end{array}\right)
\Gamma^{\mu\nu}\gamma_5\Psi^{(0)}_{\nu L}=
\frac{1}{2}\frac{\omega}{\rho}\left[e^{i\theta}\overline{\psi}^{(0)}_{\mu
L}\Gamma^{\mu\nu}\psi^{(0)*}_{\nu R}+\ h.c.\right]\, .
\label{mass}
\en

Note, however, that the above analysis in the $N=1$ case is
valid, strictly speaking, for values of
$y$ inside the semicircle, obtained from the interval $[-\pi\rho,\pi\rho]$ 
through the identification $y\leftrightarrow -y$. This leads to a discontinuity
in the transformation parameter $\cal E$
around the end-point $y=\pm\pi\rho$, since 
$U(-\pi\rho)=U^{-1}(\pi\rho)$:
\be
{\cal E}(-\pi\rho)\neq{\cal E}(\pi\rho)\, .
\label{neq}
\en
This discontinuity survives even in the large-radius limit $\rho\to\infty$ 
where the gravitino mass vanishes and supersymmetry is restored locally. This
phenomenon is reminiscent of the one found in ref.~\cite{h}, where the
discontinuity at the weakly coupled end $y=\pi\rho$ is due to the gaugino
condensate of the hidden $E_8$ formed at the strongly coupled end $y=0$. 
In fact
the two results become identical for the transformation parameter ${\cal E}$ in
the neighbourhood $y\sim\pi\rho$, in the limit $\rho\to\infty$:
\be
\lim_{\rho\to\infty}\varepsilon_L(y)\left|_{y\sim\pi\rho}\right. =
\cos\pi\omega\, \varepsilon^{(0)}_L+
\epsilon(y)\sin\pi\omega\, e^{i\theta}\varepsilon^{(0)*}_R\, .
\label{lim}
\en
On the other hand, it is easy to see that the goldstino variation vanishes in
this limit, since the discontinuity in $\partial_y\varepsilon(y)_L$ is
proportional to $\delta(y)\sin(y\omega/\rho)$.
The transformation parameter $\varepsilon_L(y)$ is thus identified 
with the spinor $\eta'$ of ref.~\cite{h}, which
solves the unbroken supersymmetry condition $\delta\psi_{5R}=0$.

Note that despite the change of 4D chirality, both terms in the r.h.s. of
eq.~(\ref{lim}) are invariant under the $\cal R$ reflection (\ref{R}), 
which defines the $N=1$ projection. 
Indeed, the second term containing the $\cal R$-odd right-handed spinor
$\varepsilon^{(0)*}_R$ is multiplied by $\epsilon(y)$, which is also odd under
$\cal R$. The proportionality constant $\sin\pi\omega$ plays the role of the
gaugino condensate in the dual description and vanishes only for
integer values of $\omega$ for which the Scherk-Schwarz mechanism becomes
trivial. In general $\omega$ is quantized, as we discussed earlier, which is
consistent with the quantization of the 
gaugino condensate through its equation of
motion that relates it with the VEV of the antisymmetric tensor
field strength
\cite{rw}. 

In the presence of gaugino condensation, the discontinuity (\ref{neq})
was interpreted as a topological obstruction 
that signals supersymmetry breaking when effects of finite radius
would be taken into account \cite{h}. 
Here we have shown that the same discontinuity,
in the infinite-radius limit, is reproduced by the Scherk-Schwarz mechanism. 
Moreover, in ref.~\cite{aq2} we have provided independent evidence 
that the finite-radius effects are also described
by the Scherk-Schwarz mechanism on the eleventh dimension, 
in the region of validity of M-theory, $\rho M_{11}\gg 1$, 
where the 10D heterotic string remains strongly coupled.

In the description of gaugino condensation by the Scherk-Schwarz mechanism, the
condensation scale is identified with the M-theory scale $M_{11}$. This implies
that the hidden $E_8$ is strongly coupled and should not contain any massless 
matter in the perturbative spectrum. Consistency then requires that the
corresponding gauge coupling be large, $\alpha_8(M_{11})\simgt 1$. On the
heterotic side, this condition follows from the minimization of the gaugino
condensation potential, which relates the value of the condensate to the
quantized VEV of the antisymmetric tensor field strength. On the M-theory side,
this provides a constraint that naively fixes the 4D unification coupling
$\alpha_G$ to be in a non-perturbative regime. Fortunately, there are important
M-theory threshold effects that invalidate this conclusion. These effects can
be understood from the lack of 
factorization of the 7-dimensional internal space
as a direct product of the semicircle with a Calabi-Yau manifold, CY$\times
S^1/{\bf Z}_2$ \cite{w}. As a result, the Calabi-Yau volume ${\widehat V}$ 
becomes a function
of $\rho$ and takes different values at the two end-points of the semicircle. 
In the large-radius limit, one finds:
\be
{\widehat V}(0)={\widehat V}(\pi\rho)-{2\pi^4}\rho M_{11}^{-3}
\left|\int_{\rm CY}{\omega\over 4\pi^2}\wedge
\left({\rm tr} F'\wedge F'-{\rm tr} F\wedge F\right)\right|\, ,
\label{thr}
\en
where $\omega$ is the K\"ahler form of CY and $F'$ $(F)$ is the field 
strength of the strongly
(weakly) coupled $E_8$ sitting at the end-point $y=0$ $(y=\pi\rho)$. 
The integral in the r.h.s. is 
a linear function of the $h_{(1,1)}-1$ K\"ahler class moduli for unit
volume, which belong to 5D vector multiplets. Its natural value is
$M_{11}^{-2}$ up to a proportionality factor of order~1~\cite{w}.

{}Following eq.~(\ref{rels}), the gauge coupling constants at the two 
end-points are related to the corresponding volumes as \cite{hw}:
\be
{1\over\alpha_G}=2M_{11}^6V(\pi\rho)\qquad ;\qquad
{1\over\alpha_8(M_{11})}=2M_{11}^6V(0)\, ,
\label{alphas}
\en
where $\alpha_G$ is the unification coupling of the weakly coupled gauge group
and the reduced volumes are defined by ${\widehat V}\equiv (2\pi)^6V$. 
Imposing now
the constraint $\alpha_8(M_{11})\simgt 1$ and using eqs.~(\ref{alphas}) and
(\ref{thr}), one finds $\rho\sim\rho_{\rm crit}$ where $\rho_{\rm crit}$
corresponds to the critical value at which the volume at the strongly coupled
end vanishes and the hidden $E_8$ decouples from the low-energy spectrum:
\ba
\rho_{\rm crit}^{-1}&=&{\alpha_G\over 16\pi^2}\, M_{11}^3
\left|\int_{\rm CY}{\omega\over 4\pi^2}\wedge
\left({\rm tr} F'\wedge F'-{\rm tr} F\wedge F\right)\right|
\nonumber\\
&\sim &{\alpha_G\over 16\pi^2}\, M_{11}\simeq 2\times 10^{-4}M_{11}
\label{crit}
\ea
Note that this condition can also be thought of as resulting
from a minimization of 
the (positive semi-definite) 4D gaugino condensation potential, which is 
proportional to $V(0)$ and, thus, vanishes at zero volume.

It is remarkable that the above relation provides the hierarchy 
necessary to fix
$\rho^{-1}$ at the intermediate scale $\sim 10^{12}$ GeV, when one identifies
the M-theory scale $M_{11}$ with the unification mass $\sim 10^{16}$ GeV
inferred by the low-energy data \cite{w,aq}. In fact, from eq.~(\ref{mass}), 
$\rho^{-1}$
determines the value of the gravitino mass and the scale of supersymmetry
breaking for the gravitational and moduli sector in the 5-dimensional bulk.
Supersymmetry breaking is then communicated to the
observable sector, living at the boundary $y=\pi\rho$, by
gravitational interactions yielding a low-energy supersymmetry breaking
$m_{\rm susy}\sim\rho^{-2}/M_p$, which is at the TeV range \cite{aq,aq2}. 
Thus, this mechanism provides an alternative ``perturbative" explanation 
of the gauge
hierarchy, where the smallness of the ratio $m_{\rm susy}/M_p\sim 10^{-16}$ 
is provided by powers of the unification coupling $\sim(\alpha_G/16\pi^2)^4$
instead of the conventional non-perturbative suppression $\sim e^{-1/\alpha_G}$.
Of course in both cases, the remaining open problem is to determine 
the actual value of the gauge coupling $\alpha_G$. 
In the present context of M-theory, this amounts to 
fixing the volume of the Calabi-Yau manifold $V(\pi\rho)$.

\section*{Note added}

After completion of this work we have received the paper of 
ref.~\cite{nilles}, 
where an independent analysis of supersymmetry breaking in M-theory is 
performed. There is however an important numerical difference in the estimate 
of the radius $\rho$ of the semicircle due to the different definition of the 
unification scale. In fact, in ref.~\cite{nilles} $M_G$ is identified with
$\widehat{V}^{-1/6}$, instead of $M_{11}$ which is also of the order of
the mass of the first KK-excitation $V^{-1/6}$  
(up to a factor of $(2\alpha_G)^{-1/6}\sim 1.5$).
As a result, our determination of $\rho^{-1}$ from the second relation of 
eq.~(\ref{rels}) yields a value which is around three orders of magnitude
below the one of ref.~\cite{nilles}. A similar difference holds for the
determination of $\rho^{-1}_{\rm  crit}$ in eq.~(\ref{crit}) for the 
same reason. Another difference concerns the gaugino masses. 
We would like to stress that the vanishing of supersymmetry breaking in
the observable sector is true only if this sector lives on a strict 4D
boundary of the 5D world. This implies that there are no KK excitations
along the eleventh (5th) dimension with quantum numbers of the Standard
Model, and thus, there are no threshold corrections to the (observable) 
gauge couplings. This reduces to a condition on the compactification
manifold implying, in particular, that the volume $V(\pi\rho)$ is
independent of $\rho$. If this requirement is not satisfied, gaugino
masses, as well as in general scalar masses, will be proportional to
the gravitino mass~\cite{largeR,modelsR}. 
Such a scenario is however phenomenologically
inconsistent since soft masses are pushed at the intermediate scale $\sim
1/\rho$ unless if the quantization condition of the Scherk-Schwarz 
parameter $\omega$ in eqs.~(\ref{mode},\ref{U}), that corresponds to the
gaugino condensate superpotential, can be avoided.

\begin{center}
{\bf Acknowledgements}
\end{center}

\noindent
We thank A.~Casas, E.~Dudas, C.~Grojean, C.~Kounnas and H.P.~Nilles for
useful discussions and comments.

\end{document}